\documentclass[12pt, english, a4paper]{scrarticle}

\usepackage{amssymb}
\usepackage{libertine}  % Good-looking modern font
\usepackage{libertinust1math}  % Accompanying math font
\usepackage[scale=0.92, varqu]{inconsolata}  % Monowidth font, varqu = upright "
\usepackage{microtype}

\usepackage[a4paper, left=3.5cm, right=3cm, top=3cm, bottom=3cm]{geometry}

\usepackage{setspace}
\usepackage{authblk}

\usepackage{graphicx}
\graphicspath{assets/}

\usepackage{array}
\usepackage[frozencache, cachedir=.]{minted}
\usepackage{booktabs}
\usepackage{multirow}
\usepackage{derivative}
\usepackage{siunitx}
\newcommand{\R}{\ensuremath{\mathbb{R}}}

\renewcommand{\hat}{\widehat}

\usepackage{xcolor}
\definecolor{MaRDIBlue}{HTML}{005eaa}      % MaRDI 100% Blue color
\definecolor{MaRDIOrange}{HTML}{d0662b}     % Slate blue for external URLs/citations

\usepackage[
  colorlinks = true,
  linkcolor=MaRDIBlue,
  citecolor=MaRDIBlue,
  urlcolor=MaRDIOrange,
]{hyperref}

\usepackage[
  natbib=false, style=numeric, giveninits=true, uniquename=init,
  isbn=false
]{biblatex}
\title{%
  Software package\\MaRDI Open Interfaces\\
  for improved interoperability\\in numerical optimization}
\author{Dmitry I.\ Kabanov, Stephan Rave, Mario Ohlberger}
\date{\small 18 June 2026}
\affil{\small \emph{Mathematics Münster, University of Münster, Germany}}

\begin{document}
\maketitle

\begin{abstract}
  To address the challenges of interoperability in computational science,
  we present the latest updates to the software package MaRDI Open Interfaces.
  This software package aims to decrease the time
  and coding/testing efforts
  spent by computational scientists on tasks
  such as writing bindings to numerical solvers
  and adapting experiment codes
  to the varying interfaces of solvers for the same problem type
  (e.g., for benchmarking, which solver is better).
  By streamlining these tasks,
  this software package helps researchers focus
  on the actual essence of their computational projects.
  Here, we demonstrate a recently developed interface for nonlinear optimization
  and illustrate how it can be applied for computational experiments
  with optimization problems.
  As an example of such problem,
  we consider training of physics-informed neural networks
  to predict the solutions of viscous Burgers' equation.
\end{abstract}

\section{Introduction}

When conducting computational experiments,
computational scientists often face two obstacles.
The first obstacle arises when a~scientist
needs to use a numerical solver written in one programming language
while their target language (language of choice) is different.
In this case, if the solver lacks bindings
to the target language,
the scientist has to develop the bindings themselves.
The second obstacle occurs, for example, in benchmarking projects
where there is a need
to switch from one solver to another
for the same type of computational problem.
Both of these situations are ``obstacles'' in a sense
that they force the scientist to spend non-negligible amount of time and energy
on coding efforts to develop the bindings,
accommodate the discrepancies in the interfaces,
and subsequently test the code changes.
Hence, the scientist must divert their attention from the core questions
of their scientific project to resolve challenging but technical aspects.

The above considerations are the basis for the primary goal of the software package
\emph{MaRDI Open Interfaces}: to increase interoperability
in computational science.
To address the first obstacle~--- different programming languages~---
this software package contains components
for automated data marshalling from one language to another.
To remove the second obstacle,
the package provides generalized (unified) interfaces
to various numerical solvers for the same type of computational problem,
such as integration of differential equations
or optimization of functions.
As a result, scientists can access solvers written in different languages
or benchmark different solvers for the same computational problem
without making substantial changes to their computational driver scripts.

In this work, we report on the recent addition
of an interface into \emph{MaRDI Open Interfaces}
that allows to solve nonlinear optimization problems.
Such problems frequently arise in computational projects,
with examples including the training of neural networks
for computer vision~\cite{Bishop2006}
or physics-related problems~\cite{RaissiEtAl2019, HeilandKim2025},
data analysis~\cite{SiviaSkilling2006},
optimization-based model order reduction~\cite{KeilOhlberger2024,
  SchwerdtnerVoigt2023, AzmiEtAl2024},
shape optimization~\cite{WyschkaWollner2025, RosandiSimeon2025}.
We therefore have added this interface
to our software package, reflecting the widespread nature of such problems
in computational science.

The paper is organized as follows.
Section~\ref{sec:landscape} provides an overview of the available
software packages for optimization problems in the languages
that are supported by \emph{MaRDI Open Interfaces} (C, Julia, and Python).
Section~\ref{sec:description} gives a brief description
of the software package and shows the interface for optimization problems.
Section~\ref{sec:results} presents the results of computational experiments
that we conduct using this interface by applying it
to the training of physics-informed neural networks.
Finally, Section~\ref{sec:conclusions} provides conclusions.

\emph{MaRDI Open Interfaces} is part of the
\emph{Mathematical Reseach Data Initiative} (MaRDI)~\cite{BennerEtAl2022},
a nationwide German Consortium for mathematical sciences
by the \emph{National Research Data Infrastructure} (NFDI).

\section{Landscape of nonlinear optimization packages}\label{sec:landscape}

In this section we give an overview of modern software packages
for nonlinear optimization in continuous variables available under open source licenses for languages such as C, Julia, and Python.

To be precise, nonlinear optimization packages solve the problems of the form
\begin{align}
  \mathrm{minimize}_{x\in\R^{n}} \quad & f(x), \quad f: \R^{n} \to \R \label{eq:minimize}\\
  \mathrm{subject\ to } \quad & g_{i}^{\mathrm L} \leq g_{i} \left( x \right) \leq g_{i}^{\mathrm U}, \quad i = 1, \dots, m^{\mathrm{constraints}}, \label{eq:constraints}\\
             & x_{j}^{\mathrm L} \leq x_{j} \leq x_{j}^{\mathrm U}, \quad j = 1, \dots, m^{\mathrm{bounds}} \label{eq:bounds},
\end{align}
where an \emph{unconstrained problem} is given only by \eqref{eq:minimize},
a~\emph{bound-constrained problem} is specified by \eqref{eq:minimize}
and \eqref{eq:bounds},
and a general \emph{constrained problem} is given by \eqref{eq:minimize}
and \eqref{eq:constraints}.

Among C libraries, \emph{NLopt} \cite{Johnson2007} is a library
implemented in C
with provided bindings to Java, Guile, Python, and Octave
providing behind a unified interface a number of
derivative-free algorithms such as COBYLA (Constrained Optimization BY
Linear Approximation) \cite{Powell1994}, BOBYQA \cite{Powell2009}, Sbplx (a reimplementation of the Subplex algorithm \cite{Rowan1990}) as well as gradient-based algorithms such as L-BFGS and truncated-Newton method
\cite{DemboSteihaug1983}.
Another solver, \emph{Ipopt} \cite{BieglerZavala2009},
is an Interior Point OPTimizer  for large scale optimization.
Written in C++, it has also bindings to C, Java, and Fortran,
as well as third-party bindings to other languages.

Python package \emph{SciPy} \cite{VirtanenEtAl2020} contains multiple solvers for nonlinear optimization,
for example,
gradient-free Nelder-Mead
\cite{NelderMead1965,GaoHan2012},
previously mentioned COBYLA,
quasi-Newton methods such as BFGS
\cite{Broyden1970a,Broyden1970b,Fletcher1970,Goldfarb1970,Shanno1970}
and its limited-memory variant L-BFGS \cite{LiuNocedal1989},
as well as variants of trust-region methods \cite{ConnEtAl2000}.
The package \emph{cyipopt} \cite{Cyipopt} provides bindings to the Ipopt
optimizer using Cython.

Julia's ecosystem also has multiple packages for nonlinear optimization.
Some of these packages are bindings to solvers in other languages
such as \emph{Ipopt.jl} that provides a wrapper to the Ipopt package
based on the JuMP
(Julia Mathematical Programming) package \cite{DunningEtAl2017},
which is a domain-specific language specifically for optimization problems
(not only continuous, but for integer variables as well).
Other packages implement optimization algorithms directly in Julia,
which allows to use the advantages of multiple dispatch,
hence, are not constrained to the floating-pointer number types
of the IEEE~754 standard \cite{IEEE754}.
To this type belongs, for example, \emph{JSOSolvers.jl} \cite{MigotEtAl2026}
that provides multiple different algorithms
for large-scale unconstrained and bound-constrained
nonlinear optimization with matrix-free approach, that is, without forming
the Hessian matrix of the objective function, particularly trust-region
Newton methods TRON~\cite{LinMore1999} and TRUNK~\cite{ConnEtAl2000}.
Another example is \emph{Optim.jl} \cite{MogensenRiseth2018}
that provides pure Julia implementations of derivative-free
algorithms such as Nelder-Mead, quasi-Newton algorithms such as
Gradient Descent and BFGS, and Hessian-required algorithms such as
Newton method with trust region and interior-point Newton method
\cite{NocedalWright2006} as well as stochastic optimization algorithms
Simulated Annealing, Adam \cite{KingmaBa2017}, AdaMax.

\section{Description of \emph{MaRDI Open Interfaces}}\label{sec:description}

Here we give a short overview description
of the package \emph{MaRDI Open Interfaces}
(for a detailed description see~\cite{KabanovEtAl2025})
and explain the interface for nonlinear optimization
that we develop for this package.

\subsection{Overview of the package}

The main principles of the architecture are:
\begin{itemize}
  \item The software is split in the user-facing and implementation-facing parts
        that are loosely coupled and communicate with each other
        via an application programming interfaces and implemented in dynamically
        loaded libraries.
  \item The users of the package communicate with the different solvers
        for the same problem type via a generalized interface.
        Correspondingly, existing solvers can be adapted
        to the generalized interface, or new solvers can be written
        directly against the interface.
  \item The performance penalty of crossing inter-language barrier
        should be as small as possible.
        This implies that the data must be passed by reference,
        not through memory copies (which are costly).
\end{itemize}

Currently, the package supports programming languages C, Julia, and Python
on both sides.
Supported data types include 32-bit signed integers, 64-bit floating-point numbers,
arrays of 64-bit floating-point numbers, strings, volatile data (opaque pointers)
and simple dictionaries (key-value pairs), and callback functions.

Arrays are always passed by reference as their size is problem-dependent,
and it would be inefficient to copy arrays on every call.
Additionally, interfaces are constructed in such a way
that the callback functions do not allocate arrays but mutate the arrays
passed as arguments, to avoid uncontrollable memory allocations.

Callback functions are implemented as a data structure that includes
a native function pointer and a function pointer to the C wrapper,
so that if the user language and the implementation language are the same,
the native function could be used;
when languages are different, then there is an overhead of
calling the C wrapper,
however, as we already demonstrated previously~\cite{KabanovEtAl2025},
the overhead is minimal and becomes negligible for larger problems,
where the useful computations take more time than data marshalling.

\subsection{Interface for nonlinear optimization}

In the current state of the project, the interface for nonlinear optimization
is able to solve unconstrained problems:
\begin{align}
  \mathrm{minimize}_{x\in\R^{n}} \quad & f(x; \alpha), \quad f: \R^{n} \times \mathcal A \to \R \label{eq:minimize-oif},
\end{align}
where the objective function \( f \) takes an additional parameter vector \( \alpha \in \mathcal A\).

Often solvers for optimization problems provide
a single function that accepts all required data:
initial guess, objective function, etc.
However, we model the optimization interface
in \emph{MaRDI Open Interfaces}
in a more object-oriented manner,
setting different required components of the problem
separately before actually solving the problem.
Such an interface allows users to catch errors earlier,
which simplifies the usage:
for example, implementations can check that an objective function
has the required signature by invoking it and reporting to the user
if there is a mismatch.
Hence, the interface
for unconstrained nonlinear optimization has currently
the following function calls available
(in Python syntax with type hinting):
\begin{minted}[escapeinside=||]{Python}
# Set an initial guess for the sought-for parameter vector |\(x\)|.
set_initial_guess(x0: numpy.ndarray)
# Set the user-provided context data for |\(f\)|.
set_user_data(user_data: Any)
# Set the callback for |\( f \)|.
set_objective_fn(objective_fn: callable)
# Set the callback for |\( \nabla_x f \)|.
set_grad_fn(grad_fn: callable)
# Set the optimization algorithm and its parameters.
set_method(method_name: str, method_params: dict)
# Minimize |\( f (x) \)| and write the resultant |\( x \)| into `out_x`.
# returning status code and a string message.
minimize(out_x: numpy.ndarray) -> typing.Tuple[int, str]
\end{minted}

The callback function \texttt{objective\_fn} and \texttt{grad\_fn}
has the following signatures
\begin{minted}[escapeinside=||]{Python}
f(x: numpy.ndarray, user_data: typing.Any) -> float
g(x: numpy.ndarray, grad_values: numpy.ndarray, user_data: Any) -> None
\end{minted}
where the argument \texttt{grad\_values} is mutable.

In the current version of \emph{MaRDI Open Interfaces},
the following adapters for existing optimization packages are available:
\begin{itemize}
  \item The \texttt{scipy\_optimize} adapter
        for the \texttt{optimize} subpackage of
        \emph{SciPy}~\cite{VirtanenEtAl2020},
  \item The \texttt{optim\_jl} adapter
        for the \emph{Optim.jl} package~\cite{MogensenRiseth2018}.
\end{itemize}

\section{Results}\label{sec:results}

In this section we conduct computational experiments
using the interface for nonlinear optimization
to train a physics-informed neural network~\cite{RaissiEtAl2019}
as a surrogate model for a solution of a PDE problem.

\paragraph{Problem formulation}
As a PDE problem we consider the viscous Burgers' equation~\cite{Burgers1948}
with a smooth initial and periodic boundary conditions:
\begin{align}
  \begin{aligned}
\label{eq:vbe}
  \pdv{u}{t} + u \pdv{u}{x} &= \epsilon \pdv[order=2]{u}{x}, \\
          u(0, x) &= 0.5 - 0.25 \sin(\pi x), \\
          u(t, 0) &= u(t, 2),
  \end{aligned}
\end{align}
where solution \(u: \R \times \R \to \R \)
with time domain \(t \in [0, 2]\), spatial domain \(x \in [0, 2]\),
and parameter \(\epsilon\). We take \(\epsilon = 0.01 / \pi \) here.

For the PDE problem~\eqref{eq:vbe} we would like to train a model
\(\hat u(\theta; t, x)\)
to work as a predictor of the solution of the problem.

The unknown parameter vector \( \theta \) is found
via the training process (that is, optimization) based on enforcing
the PDE residual, the initial condition, and the boundary condition
on a set of collocation points in the space-time domain of the problem~\eqref{eq:vbe}
replacing the exact solution \( u(t, x) \) with its approximation \( \hat u(\theta; t, x) \).
Collocation points are created by covering the time-space domain
of the problem ~\eqref{eq:vbe} with uniform grids in time and space.
We denote the resultant grid points
as sets \(\mathbb T\) and \(\mathbb X\)
with their cardinalities being \(N_t\) and \(N_x\), respectively.

Training process is thus to find a minimizer (optimal vector \(\theta\)) of the loss function:
\begin{align}
  \label{eq:loss}
  \mathcal L(\theta; \mathbb T, \mathbb X) = \mathcal L_{\mathrm{eq}} + \mathcal L_{\mathrm{ic}} + \mathcal L_{\mathrm{bc}},
\end{align}
where the loss function \(\mathcal L_{\mathrm{eq}} \) is the loss with respect to the PDE:
\begin{align}
  \label{eq:loss-eq}
  \mathcal L_{\mathrm{eq}}(\theta; \mathbb T, \mathbb X) =
    \frac{1}{N_tN_x} \sum_{i=1}^{N_t} \sum_{j=1}^{N_x}\left(
      \pdv{\hat u(\theta; t_i,x_j)}{t} +
      \hat u(\theta; t_i,x_j) \pdv{\hat u(\theta; t_i,x_j)}{x} -
      \epsilon \pdv[order=2]{\hat u(\theta; t_i, x_j)}{x}
    \right)^2,
\end{align}
loss function \( \mathcal L_{\mathrm{ic}} \) defines
how well the initial condition is satisfied:
\begin{align}
  \mathcal L_{\mathrm{ic}}(\theta; \mathbb X) = \frac1{N_x} \sum_{j=1}^{N_x} \left(
      0.5 - 0.25 \sin(\pi x_j) - \hat u (\theta; 0, x_j)
  \right)^2,
\end{align}
and the loss function \( \mathcal L_{\mathrm{bc}} \) defines
how well the periodic boundary condition is satisfied:
\begin{align}
  \mathcal L_{\mathrm{bc}}(\theta; \mathbb T) = \frac1{N_t} \sum_{i=1}^{N_t} \left(
    \hat u(\theta; t_i, 2) - \hat u(\theta; t_i,0)
  \right)^2.
\end{align}

For the approximation model \(\hat u(\theta; t, x)\),
we choose a neural network
of the multilayer perceptron type~\cite{Bishop2006}:
\begin{align}
  \label{eq:def-mlp}
\hat u (\cdot; t, x) = f_{L} \circ f_{L-1} \dots \circ f_1,
\end{align}
with functions \( f_\ell \), \( \ell = 1, \dots, L \), being defined as
\begin{align}
f_\ell(z_{\ell-1}) &= h_\ell^\odot(W_\ell z_{\ell-1} + b_\ell), \quad \ell = 1,\dots,L-1,\\
f_L(z_{L-1})      &= W_L z_{L-1} + b_L,
\end{align}
where \(z_0 = (t, x)\), functions \(h_\ell^\odot(\cdot) = \tanh \cdot  \)
are activation functions applied componentwise to their input vectors,
and matrices \(W_\ell\) and vectors \(b_\ell\), \( \ell = 1, \dots, L \), are unknown parameters.
The sought-for vector \(\theta\) is a vectorization of these matrices and vectors.

\paragraph{Experiment setup}
We solve the minimization problem for the loss function~\eqref{eq:loss} using the BFGS
method~\cite{Broyden1970a,Broyden1970b,Fletcher1970,Goldfarb1970,Shanno1970},
which is a gradient-based quasi-Newton method,
with the implementations from the \emph{SciPy} and \emph{Optim.jl} packages.
The BFGS implementation in \emph{Optim.jl} can be used
with different line-search strategies, and we use three of them:
\texttt{HagerZhang}
\cite{HagerZhang2006},
\texttt{StrongWolfe}, and \texttt{BackTracking};
the last two algorithms are based on Chapter~3 of~\cite{NocedalWright2006}.
On the other hand, the BFGS method in \emph{SciPy} is coupled
to the line-search strategy based on strong Wolfe conditions.
This amounts to effectively four different implementations
that we consider.

To compute the gradient of the loss function~\eqref{eq:loss}
with respect to the vector \(\theta\)
as well as to compute the derivatives of \(\hat u\) in~\eqref{eq:loss-eq},
we use automatic differentiation.
For that reason, the code is implemented in Python
using the JAX library~\cite{JAX2018Github}
that provides the required functionality.
As a remark, we would like to point out that it was crucial
to provide a callback for the gradient function for the training process,
as using finite-difference approximations of the gradient was not sufficient for successful training.
We set the default data type for JAX to be 64-bit floating-point numbers,
and run computations only on CPU by setting \texttt{JAX\_PLATFORMS=cpu};
the reported runtimes below are thus obtained on Intel Xeon processors.

The stopping criterion requires the \(\infty\)-norm of the gradient vector
to be less than \(\texttt{gtol}\),
for which we take values from \( \{\num{1e-3}, \num{1e-4}, \num{1e-5} \} \).

The architecture of the neural network \(\hat u\)
is instantiated with the configuration \(\{2, 21, 13, 8, 5, 3, 2, 1\}\),
where the first (input) and the last (output) values are fixed.
With this configuration, the sought-for parameter vector \(\theta \in \R^{535}\).

The optimization process requires an initial guess on the value of \(\theta\),
which is initialized using the Xavier initialization~\cite{GlorotBengio2010},
All implementations start the optimization process from the same initial guess
(the seed for the random number generator).

To assess the influence of the initial guess on the solution,
number of iterations, and so on, we conduct 42 independent trials
where the seed of the random number generator is different in each trial
and equal to the number of the trial.
Correspondingly, all measured statistics are reported below
as the mean values over these 42 trials, with uncertainties being
standard errors of the mean values~\cite{Wasserman2004}.

\paragraph{Experiment results}
The results of the experiment are summarized in~\autoref{tab:vbe-bfgs},
where different rows correspond to the pairs of an implementation
and the stopping tolerance of the gradient vector
and the columns correspond to the different statistics of optimization
process: number of iterations,
number of the loss function and its gradient evaluations,
attained final gradient-norm and loss values, and the elapsed time in seconds.

Considering the number of iterations, \emph{SciPy} results
are close to the results achieved by the pair \emph{Optim.jl}-\texttt{StrongWolfe}
as they use the same line-search strategy.
Implementation \emph{Optim.jl} with \texttt{HagerZhang}
(the default line search)
performs comparably to the pair \emph{Optim.jl}-\texttt{StrongWolfe}.
For tighter tolerances, \emph{Optim.jl} with \texttt{BackTracking}
is slightly more efficient for this problem, although the \texttt{BackTracking}
line search is not as strict as \texttt{StrongWolfe}.

Looking at the column \texttt{nfuncevals}
(function and gradient evaluations  have the same values),
we can see that the pairs \emph{Optim.jl}-\texttt{HagerZhang}
and \emph{Optim.jl}-\texttt{StrongWolfe} require the most number
of function and gradient evaluations,
while for \emph{SciPy} and \emph{Optim.jl}-\texttt{BackTracking}
the number of evaluations is at most 10\% larger than the number of iterations
(except for the case \( \texttt{gtol} = \num{1e-3} \) for the latter implementation).

From the columns \texttt{final\_loss},
one can see that all implementations arrive at approximately the same
values of the final loss values,
which indicates that the loss surface is relatively regular.

The last column of~\autoref{tab:vbe-bfgs} provides the elapsed times
in seconds,
which we measured ourselves using the \texttt{perf\_count()} Python function
(the algorithms in \emph{Optim.jl} also report the elapsed times,
however, they are systematically lower than the ones measured by us).
Probably, the most interesting in this column is that
despite of crossing the language boundaries,
the implementation \emph{Optim.jl}-\texttt{BackTracking}
outperforms on average the \emph{SciPy} implementation,
even when the number of iterations for \emph{SciPy} is smaller.
This demonstrates that \emph{MaRDI Open Interfaces} provide
the flexibility of using different solvers in different languages
without incurring performance penalties.

Overall, one can see, especially comparing implementations \emph{SciPy}
and \emph{Optim.jl}-\texttt{StrongWolfe}, that although they have
theoretically the same algorithms for the optimization process
and for the line-search strategy,
the performance in terms of the number of function evaluations is not the same:
for example, for the case \( \texttt{gtol} = \num{1e-5} \)
the average number of function evaluations differ by 17\%.
Both implementations state that they use the algorithms from Chapter 3
of the book by Nocedal \&~Wright~\cite{NocedalWright2006} for line search
based on the strong Wolfe conditions: the sufficient decrease condition
that ensures that the step is not too large and the curvature condition
that ensures that the step is not too small.
One explanation to the discrepancies in performance is
that the line-search strategy in \emph{SciPy},
at stated in the source-code documentation,
computes the second Wolfe condition (the curvature condition)
only when the first condition (sufficient decrease)
fails to yield a suitable step size, while \emph{Optim.jl}-\texttt{StrongWolfe}
seems to compute both conditions every time.
This behavior shows that even when different implementations implement
the same algorithms from the same articles and books,
their actual real life performance can vary significantly.

\begin{table}[htbp]
\centering
\caption{
  Statistics on training a physics-informed neural network~\eqref{eq:def-mlp}
  on the problem~\eqref{eq:loss} using implementations of the BFGS method
  from \emph{SciPy} and from \emph{Optim.jl}
  with different line-search strategies
  via \emph{MaRDI Open Interfaces}
  with different values of the tolerance on the \(\infty\)-norm
  of the gradient vector \texttt{gtol}.
  All reported statistics are mean values over 42 trials,
  with each trial using the same initial guess
  for the parameters of a neural network
  for all implementations.
  Uncertainties are standard errors of the mean.\label{tab:vbe-bfgs}
}
\newcommand{\RowSciPy}{\multirow[c]{3}{*}{SciPy}}
\newcommand{\RowOptimHZ}{\multirow[c]{3}{*}{\parbox{2cm}{Optim.jl\\HagerZhang}}}
\newcommand{\RowOptimSW}{\multirow[c]{3}{*}{\parbox{2cm}{Optim.jl\\StrongWolfe}}}
\newcommand{\RowOptimBT}{\multirow[c]{3}{*}{\parbox{2cm}{Optim.jl\\BackTracking}}}
\sisetup{output-exponent-marker = \mathrm{e}}
\begin{tabular}{l l
  S[table-format = 3.1 \pm 2.1]  % N iterations
  S[table-format = 3.1 \pm 2.1]  % N function evaluations
  S[table-format = 1.1e-1]       % Final loss
  S[table-format = 3.1 \pm 2.1]  % Runtime, seconds
}
\toprule
            &        & {niterations}  & {nfuncevals}  & {final\_loss}       & {elapsed, sec.} \\
{impl}      & {gtol} &                &               &                     &                 \\
\midrule
\RowSciPy   & 1e-3   &  74.0 +-  3.8  &  80.1 +-  3.9 & \num{3.0 +- 0.7e-3} &  22.6 +-  0.7   \\
            & 1e-4   & 278.0 +-  9.7  & 288.1 +- 10.0 & \num{9.5 +- 0.3e-5} &  59.0 +-  1.8   \\
            & 1e-5   & 672.2 +- 28.7  & 690.8 +- 29.5 & \num{5.9 +- 0.3e-6} & 129.0 +-  5.2   \\
\midrule
\RowOptimHZ & 1e-3   &  65.2 +-  4.0  &  90.2 +-  5.5 & \num{5.7 +- 1.3e-3} &  22.4 +-  0.9   \\
            & 1e-4   & 308.0 +- 12.2  & 355.4 +- 14.5 & \num{6.5 +- 0.4e-5} &  66.0 +-  2.4   \\
            & 1e-5   & 677.8 +- 29.0  & 753.9 +- 34.0 & \num{6.7 +- 0.6e-6} & 131.1 +-  5.5   \\
\midrule
\RowOptimSW & 1e-3   &  70.2 +-  3.2  & 107.9 +-  5.6 & \num{3.1 +- 0.7e-3} &  23.1 +-  0.8   \\
            & 1e-4   & 278.3 +- 12.8  & 351.9 +- 20.5 & \num{1.3 +- 0.6e-4} &  61.5 +-  3.0   \\
            & 1e-5   & 677.7 +- 60.3  & 814.2 +- 82.2 & \num{6.3 +- 0.5e-6} & 133.9 +- 12.2   \\
\midrule
\RowOptimBT & 1e-3   &  77.8 +-  4.0  &  95.3 +-  5.2 & \num{2.9 +- 0.7e-3} &  20.4 +-  0.7   \\
            & 1e-4   & 268.9 +-  8.9  & 296.7 +- 10.7 & \num{7.2 +- 0.5e-5} &  52.4 +-  1.6   \\
            & 1e-5   & 629.6 +- 30.7  & 676.9 +- 34.5 & \num{6.1 +- 0.4e-6} & 112.3 +-  5.1   \\
\bottomrule
\end{tabular}
\end{table}

\begin{figure}[tbp]
  \centering
  \includegraphics[width=\textwidth]{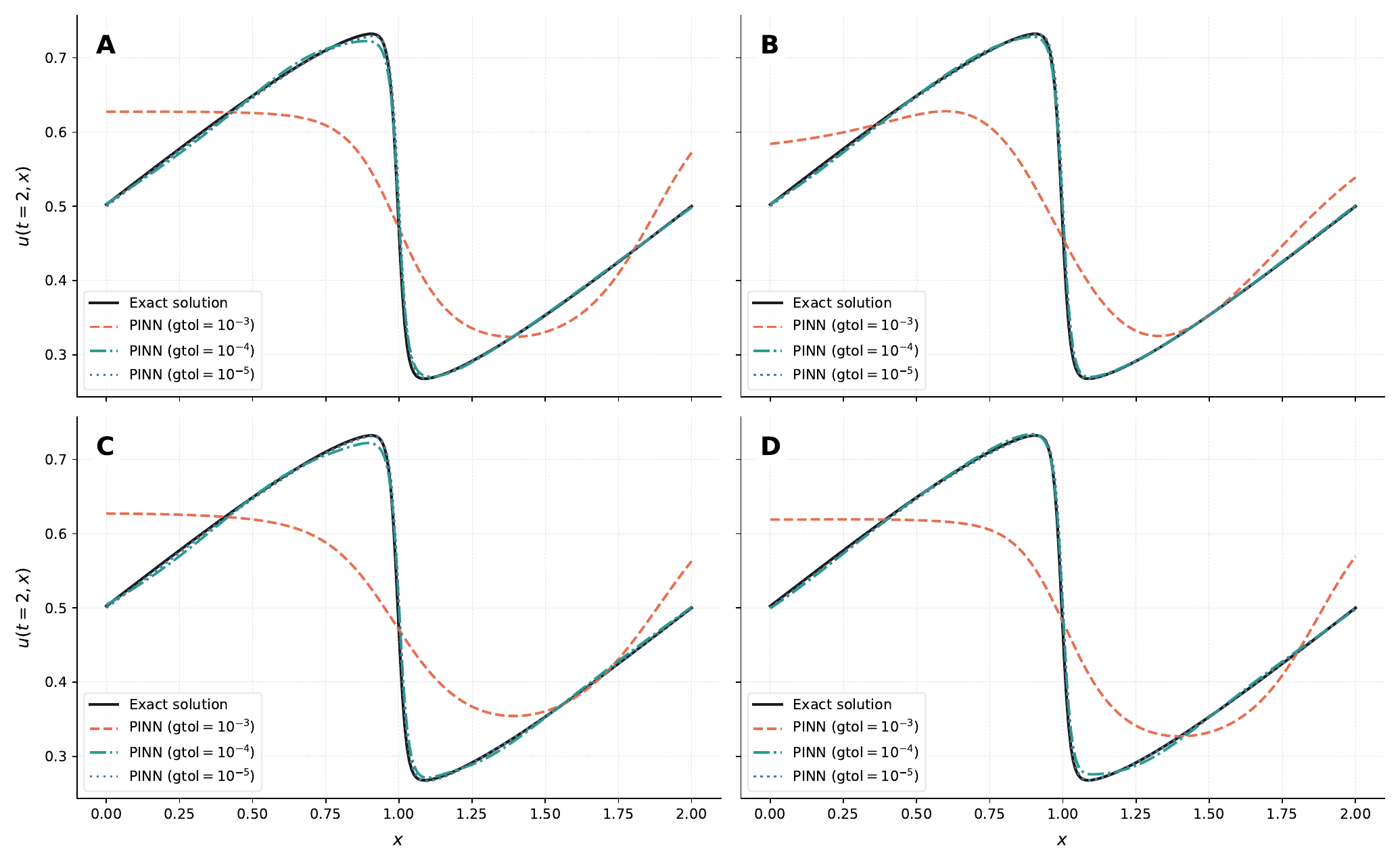}
  \caption{Comparison of the solutions
    of the problem~\eqref{eq:vbe} at \(t = 2\):
    exact solution via the Cole--Hopf transformation versus
    approximated solutions by physics-informed neural networks~\eqref{eq:def-mlp}
    trained until the gradient \(\infty\)-norm tolerance \(\texttt{gtol}\) is met
    for \(\texttt{gtol} \in \{\num{1e-3}, \num{1e-4}, \num{1e-5}\}\)
    via the BFGS method for trial 1.
    \textbf{A}~Using \emph{SciPy}
    \textbf{B}~Using \emph{Optim.jl}-\texttt{HagerZhang}
    \textbf{C}~Using \emph{Optim.jl}-\texttt{StrongWolfe}
    \textbf{D}~Using \emph{Optim.jl}-\texttt{BackTracking}\label{fig:vbe-bfgs}.}
\end{figure}

\autoref{fig:vbe-bfgs} shows the solutions,
predicted by trained neural networks
for different values of \texttt{gtol} against the exact solution at \(t = 2\)
from trial 1:
each subfigure corresponds to the different implementation.
The exact solution of the problem in all subfigures
is computed using the Cole--Hopf transformation\cite{SalsaVerzini2022}
and solving the resultant problem for the heat equation using Fourier series.
As one can see, the approximation of the solution by the neural network
is in a very good agreement with the exact solution
when the training process continues
until the stopping criterion \(\texttt{gtol} \leq \num{1e-5}\) is met.
For \( \texttt{gtol} = \num{1e-3} \)
all implementations predict the solution rather crudely,
particularly, the periodic boundary condition is not satisfied at all.
Analysis of other trials
(hence, other initial guesses for the neural-network parameters)
shows that for some initial guesses even for \( \texttt{gtol} = \num{1e-4} \),
the predicted solutions can be unsatisfactory (violating the boundary condition).
However, tighter tolerance \( \texttt{gtol} = {1e-5} \)
allows to obtain a neural network that predicts the solution in good agreement
with the exact solution independent from the used implementation
and the line-search strategy as well as from the initial guess.

\section{Conclusions}\label{sec:conclusions}

We have demonstrated the new open interface
for nonlinear optimization problems added to the software package
\emph{MaRDI Open Interfaces}.
Although work-in-progress~--- only unconstrained
optimization problems are supported for now~--- the interface provides
functionality for conducting computational experiments with optimizers
available in the \emph{SciPy} and \emph{Optim.jl} packages.

As the results of computational experiments with training a physics-informed
neural network for the viscous Burgers' equation demonstrate,
even when the problem is implemented in one programming language,
one can use an optimizer from another language without significant
performance overhead. Comparison of different implementations
shows that even when they implement the same algorithms
as is the case for the implementation from \emph{SciPy}
and the \emph{Optim.jl}-\texttt{StrongWolfe} pair, the actual real life
performance in terms of the number of iterations and other statistics
of the solution process can vary widely.

Ability to benchmark numerical solvers for the same problem type
minimizing the number of changes such as rewriting the loss function
in a different programming language is an important property
of \emph{MaRDI Open Interfaces} that can help improve interoperability
in scientific computing.
Besides that, the coding and testing efforts on switching between
different solvers are accommodated inside the \emph{MaRDI Open Interfaces}
relieving computational scientists from spending time and energy on these tasks
and keeping their focus on the core scientific questions.

\begin{small}\paragraph{Acknowledgements}
This work was funded by Nationale Forschungsdaten Infrastruktur
(NFDI, National Research Data Infrastructure),
project number~460135501, NFDI~29/1
“MaRDI – Mathematical Research Data Initiative
[Mathematische Forschungsdateninitiative]”
and~additionally by the Deutsche Forschungsgemeinschaft (DFG, German Research Foundation) under Germany's Excellence Strategy EXC 2044/2 - 390685587, Mathematics Münster: Dynamics-Geometry-Structure.
\end{small}

\begin{small}\paragraph{Conflict of interests}
    The authors declare no conflict of interests.
\end{small}

\printbibliography
\end{document}